\documentclass[aps,prb,preprint,graphicx,amsbsy,amssymb]{revtex4}

\usepackage{graphicx}
\usepackage{graphics}
\usepackage{epsfig}

\begin{document}
\draft
\title{ Ginzburg-Landau equation and vortex liquid phase of Fermi liquid superconductors}
\author{Tai-Kai Ng and Wai-Tak Tse}
\address{
Department of Physics,
Hong Kong University of Science and Technology,
Clear Water Bay Road,
Kowloon, Hong Kong
}
\date{ \today }

\begin{abstract}
  In this paper we study the Ginzburg-Landau (GL) equation for Fermi liquid superconductors with strong Landau
 interactions $F_0$ and $F_1$. We show that Landau interactions renormalize two parameters entering the GL
 equation leading to renormalization of superfluid density and separation of length and energy scales for amplitude
 and phase fluctuations. The separation of energy/length scales for amplitude and phase fluctuations in term leads
 to unconventional (2D) KT transition and vortex liquid phase. Application of the GL equation to describe underdoped
 high-$T_c$ cuprates is being discussed.
\end{abstract}


 \maketitle


    The understanding of the physics behind high-$T_c$ cuprates remains one of the major challenges to the
 condensed matter physics community nowadays\cite{r1,r2,r3}. One big mystery is the pseudo-gap phase in underdoped
 cuprates where a d-wave like gap in the quasi-particle spectrum exists, but the system is already a normal
 metal\cite{r1,r2,r3}. It has been suggested that the pseudo-gap phase can be understood as a vortex flow state of a
 layered superconductor with strong phase fluctuations\cite{emery} existing in temperature range $T_c<T<T^*$ ($T_c=$
 superconductor transition temperature and $T^*=$ pseudo-gap temperature). To test this idea, the Ginzburg-Landau
 (GL) action for a strong coupling superconductor corresponding to the physical picture of preformed pairs has been
 derived and studied\cite{rand,levin}. However the GL action is not able to explain the different trends of $T_c$ and
 $T^*$ as functions of doping $x$ in the pseudo-gap phase, presumably due to the lack of consideration of the effect
 of proximity to Mott transition\cite{r2,r3,emery}. In this paper we shall explore this problem further by including the
 effect of Landau interactions in the GL action. We shall derive the GL action for a Fermi-liquid superconductor
 with strong density-density and (transverse) current-current Landau interactions $F_0$ and $F_1$. We note that the
 GL action we derive is applicable to general Fermi liquid superconductors with strong Landau interactions $F_0$ and
 $F_1$ and is not restricted to High-$T_c$ cuprates . We shall show how Landau interactions renormalize two parameters
 entering the GL action leading to renormalization of superfluid density and separation of length and energy scales
 for the amplitude and phase fluctuations. The effects of Landau interaction on superfluid dynamics, including
 critical magnetic fields and critical current, vortex structures and vortex liquid phase, will be examined. We find
 that Landau interactions provide a natural mechanism for the separation of temperature scales $T^*$ and $T_c$ in the
 underdoped cuprates. However many properties of the pseudo-gap phase remain unexplained\cite{r2,r3,e1}.

   In Fermi liquid theory, the long-wavelength and low frequency electromagnetic response of a Fermi liquid
 superconductor to external electromagnetic perturbations can be described by an effective Hamiltonian\cite{leggett}
 \begin{equation}
 \label{hmfs}
 H=H^{BCS}+{1\over2}\sum_{q}N_F^{-1}\left({F_1\over k_F^2}\vec{j}_{t}(q).\vec{j}_t(-q)+F_0n(q)n(-q)\right)
 \end{equation}
 where $H^{BCS}$ is the mean-field BCS-Hamiltonian, $q=(\vec{q},\omega)$. $N_F$ is the density of state on the
 Fermi surface, $k_F$ is the Fermi momentum and $F_1$ and $F_0$ are the Landau parameters describing (transverse)
 current-current and density-density interactions in the system, respectively. We shall not restrict ourselves to
 translational invariant systems here and therefore there is no particular relation between the Landau parameter
 $F_1$ and effective mass $m^*/m$. We disregard all other Landau interactions in this paper since their effects on
 the GL action are much weaker. Integrating out the fermion fields, the (transverse) current and density responses
 of the system to EM field is given by the effective action\cite{leggett}

 \begin{eqnarray}
 \label{leff}
 S_{eff} & = &
 \sum_{q}\left({1\over2}\left({1\over K_0(q;T)}-{F_{1}\over(1+F_{1})\bar{K}_0(0)}\right)\vec{j}_t^2+
 {1\over2}\left({1\over\chi_0(q;T)}-{F_0\over\bar{\chi}_0(0)}\right)n^2-\vec{j}_t.\vec{A}-n\phi\right)   \\
\nonumber
  & = &  \sum_{q}\left({1\over2}\left({\vec{j}_t^2\over K_0(q;T)}+{n^2\over\chi_0(q;T)}\right)
  -\vec{j}_t.(\vec{A}+\vec{a})-n(\phi+\varphi)+{1\over2}\left((1+F_{1}^{-1})\bar{K}_0(0)\vec{a}^2+{\bar{\chi}_0(0)
  \over F_0}\varphi^2\right)\right)
 \end{eqnarray}
 where $T=$ temperature and $K_0(q;T)$ and $\chi_0(q;T)$ are the transverse current-current and density-density
 response functions for the BCS superconductor in the absence of Landau interactions, respectively. $-\bar{K}_0(T)
 =-K_0(\vec{q}\rightarrow0,\omega=0;T)$ and $-\bar{\chi}_0(T)=-\chi_0(\vec{q}\rightarrow0,\omega=0;T)$ are the
 corresponding superfluid density and compressibility of the BCS superconductor. $\vec{A}$ and $\phi$ are the
 external electromagnetic vector and scalar fields. Fictitious gauge potentials $\vec{a}$ and $\varphi$
 are introduced to decouple the current-current and density-density interactions (Legendre transformation) in the
 second line of Eq.\ (\ref{leff}). In this representation, the Landau interactions are absorbed by introducing
 fictitious vector and scalar fields that couples to the BCS superconductor.

  The corresponding Ginzburg Landau action is therefore of the form
  \begin{eqnarray}
  \label{gleff}
  S_{GL} & = & \int d^dx\int
  dt\left(-i\gamma\psi^*(\partial_t+ie^*(\phi+\varphi))\psi+
  +{1\over2}{\bar{\chi}_0(0)\over F_0}\varphi^2\right.    \\ \nonumber
  & & +\left.{\hbar^2\over2m^*}|(\nabla-i{e^*\over \hbar c}(\vec{A}+\vec{a})\psi|^2-\alpha(T)|\psi|^2+
  {\beta\over2}|\psi|^4+{1\over2}{\bar{K}_0(0)\over G}\vec{a}^2\right),
  \end{eqnarray}
  where $G=F_1/(1+F_1)$, which is the GL action for a BCS superconductor coupling to the effective electromagnetic
  fields $\phi+\varphi$ and $\vec{A}+\vec{a}$. For weak-coupling BCS superconductors, $\alpha(T)\sim\epsilon(T^2_M/E_f)$,
  where $\epsilon=1-T/T_M$, $\beta\sim N_F^{-1}(T_M/E_f)^2$, and $\gamma=\gamma^{'}+i\gamma^{"}$, where
  $\gamma^{'}\sim(T_M/E_f)^2$, and $\gamma^{"}\sim T_M/E_f$. $T_M$ is the mean-field (BCS) transition temperature and
  $E_f$ is the Fermi energy\cite{rand}. In the strong coupling limit, the GL action becomes the Gross-Pitaevski
  action for a gas of charge $e^*=2e$ bosons, with $\alpha(T)\rightarrow\bar{\mu}\sim-\epsilon E_b/2$ being the
  chemical potential for the bosons, where $E_b\sim T_M$ is the bound state energy for the electron pair,
  $m^*\rightarrow2m$, $\gamma\rightarrow1$, and $\beta\rightarrow4\pi a_b/m^*$, where $a_b>0$ is the (composite)
  boson scattering length\cite{rand}. Consistency between Eq.\ (\ref{leff}) and Eq.\ (\ref{gleff}) implies that
  $\bar{K}_0(T)=-(e^{*2}\alpha)/(m^*c^2\beta)$ and $\bar{\chi}_0(T)=-(\gamma_1e^*)^2/\beta$.

   The fictitious gauge fields $\vec{a}$ and $\varphi$ can be eliminated easily from Eq.\ (\ref{gleff}). Writing
   $\psi=\sqrt{\rho}e^{i\theta}$, we obtain
  \begin{equation}
  \label{gleff2}
  S_{GL}\rightarrow\int d^dx\int dt\left(\rho\gamma(\partial_t\theta+e^*\phi)
  +{\hbar^2\over2m^*}(\nabla\sqrt{\rho})^2-\alpha(T)\rho+{\bar{\beta}\over2}\rho^2
    +{\hbar^2\over2m^*}{\rho\over(1+G'(\rho))}(\nabla\theta-{2\pi\over\Phi_0}\vec{A})^2\right).
  \end{equation}
  where $\Phi_0=hc/e^*$ is the fluxoid quantum, $\bar{\beta}=\beta(1+F_0)$ and
  $G'(\rho)=-G\rho(T)/\rho(0)$, where $\rho(0)=\alpha(T=0)/\beta$.

  Equation\ (\ref{gleff2}) is the main result in this paper. We observe that the Landau interactions renormalize
  two parameters in the GL action, with $\beta\rightarrow\bar{\beta}=\beta(1+F_0)$ and the superfluid density
  $\rho_s$ (or London penetration depth $\lambda^{-2}$) renormalized to
  \begin{equation}
  \label{sdensity}
  \rho_s(T)={\rho(T)\over(1+G'(\rho))}={(1+F_1)\rho(T)\over1+F_1(1-\rho(T)/\rho(0))},
  \end{equation}
  in agreement with result from Fermi liquid theory\cite{leggett}.

  We now discuss a few general consequences of the renormalized GL action. We shall concentrate on the
  renormalization effects associated with $F_1$ since $F_0$ can be absorbed simply by putting
  $\beta\rightarrow\bar{\beta}$. The transition temperature $T_c$ (governed by $\alpha(T)$) and characteristic
  (coherence) length $\xi^2_A(T)=\hbar^2/(2m^*\alpha(T))$ governing the gap amplitude fluctuations are not
  renormalized by Landau interactions. However the coherence length governing phase fluctuation is renormalized, with
  $\xi_P(T)\sim\xi_A(T)/\sqrt{(1+G'(\rho))}$. The two length scales separate in the presence of Landau interaction
  $F_1$. Notice that the renormalization effects associated with $F_1$ is proportional to $\rho(T)$. Therefore, at
  $T\sim T_M$ or $H\rightarrow H_{c2}$ where $\rho(T)\rightarrow0$, the renormalization effects of $F_1$ are not
  important and the only effect of Landau interaction is renormalization of $\beta$.

   Next we consider critical magnetic fields and critical currents. The thermal
   dynamical critical field $H_c$ and the upper critical field $H_{c2}\sim\Phi_0/2\pi\xi^2$ are not renormalized by
   $F_1$. The later is because $\rho\rightarrow0$ at $H\rightarrow H_{c2}$, and $F_1$ becomes unimportant. The lower
   critical field $H_{c1}\sim(\Phi_0/4\pi\lambda^2)ln(\lambda/\xi_P)$ is renormalized by $F_1$ through both
   $\lambda=\sqrt{(1+G'(\rho))}\lambda_0$ and $\xi_P\sim\xi_A/\sqrt{(1+G'(\rho))}$, where $\lambda_0$ is the
   London penetration depth of the corresponding pure BCS superconductor and $\xi_P$ is the effective vortex core
   size defined by super-currents (see below). In particular, for $1+F_1<1$ $\lambda>\lambda_0$, $\xi_P<\xi_A$ and
   $H_{c1}$ is reduced by $F_1$.

   The critical current passing through thin wire or film can be determined by minimizing the free energy with fixed
   velocity $\vec{v}_s=\hbar(\nabla\theta-{2\pi\over\Phi_0}\vec{A})$ and then maximizing the supercurrent
   $\vec{j}_s=e^*\rho_s\vec{v}_s$\cite{tinkham}. It is easy to see that the critical current is reduced (enhanced) by nonzero
   $F_1<(>) 0$, and the reduction (enhancement) is stronger at lower temperature. As a result, the rate of reduction
   of the critical current as temperature increases is slower (faster) than the corresponding BCS superconductor
   when $F_1<(>) 0$. The rate of decrease can be fit roughly by the formula $j_c(T)\sim j_c(0)(1-T/T_M)^{\nu}$,
   where $\nu=3/2$ for BCS superconductors and changes continuously when $F_1$ changes. We find numerically that
   $\nu\sim1.35$ for $F_1=-0.9$ and increases to $\nu\sim1.7$ for $F_1=0.5$.

      Next we discuss vortices. First we consider single vortex solution of the GL equation, i.e. solution of form
   $\psi(r,\phi)=f(r)e^{i\phi}$. We shall consider $T=0$ for simplicity. The separation of length scale associated
   with amplitude and phase fluctuations implies that two "sizes" of the vortex core can be defined, one $\sim\xi_A$
   is the size of region where the amplitude of the BCS wavefunction $f(r)$ goes to zero. This is not renormalized by
   Landau interaction. The other $\sim\xi_P$ is the size of region where the superfluid density $\rho_s(r)$ defined
   by Eq.\ (\ref{sdensity}) goes to zero. The two coherence lengths differs when Landau interaction $F_1$ is nonzero.
   To see that $\xi_P$ represents the core size defined by the superfluid density, one may replace the density
   variable $\rho$ by the superfluid density variable $\rho_s$ in the GL action using Eq.\ (\ref{sdensity}) and
   derive the corresponding GL equation for $\rho_s$ and $\theta$. Performing a small $r$ expansion for
   the vortex solution ($\nabla\theta\sim\hat{\phi}/r$) It is straightforward to see that $\xi_P$ represents
   the coherence length for superfluid density. In figure 1 we show $\rho_s(r)/\rho_s(r\rightarrow\infty)$ as
   a function of $r/\xi_A$ at zero temperature solved numerically for three different values of $F_1=-0.5,0,1$. The
   dependence of superfluid density vortex core size on $F_1$ is clear.
   \begin{figure}
   \includegraphics[width=6.0cm, angle=0]{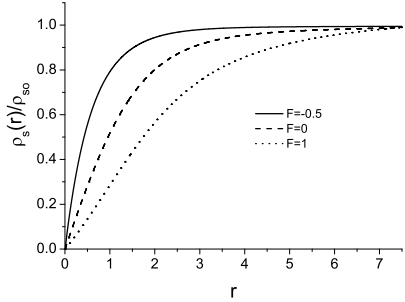}
   \caption{\label{Fig.1} superfluid density $\rho_s(r)/\rho_{so}$ $(\rho_{so}=\rho_s(r\rightarrow\infty))$ as function of $r$ for three values
   of $F_1=1.0,0.0,-0.5$.}
   \end{figure}

     Vortex viscosity is also affected by Landau interactions. In standard Bardeen-Stephen (BS) type
     arguments\cite{tinkham}, the vortex flow viscosity $\eta_v$ is given by
     \[
     \eta_v\sim({a_v\over\xi^2})\sigma_L,   \]
     where $a_v$ is a constant of order O(1) and $\xi,\sigma_L$ are the effective vortex core size and normal state
   conductivity for the Fermi liquid superconductor, respectively. Both $\xi$ and $\sigma_L$ are renormalized by
   $F_1$, with $\sigma_L\sim(1+F_1)^2\sigma_0$ where $\sigma_0$ is the conductivity for the normal metal
   without Landau interactions and $\xi\sim\xi_P$.

     The renormalized GL action provides a natural explanation to one problem faced by the ordinary GL
   action (in both the weak or strong coupling limit) when applying to underdoped High-$T_c$ cuprates - the
   vanishing of superfluid density $\rho_s$ when concentration of hole $x\rightarrow0$ and the huge difference
   between the temperature $T_c$ and $T^*$ in the pseudo-gap phase. In models of strong correlations, the vanishing
   of superfluid density is a direct consequence of proximity to Mott Insulator and is reflected in the Landau
   parameters $F_1$, with $F_1\sim x-1$ at $T\rightarrow0$ in gauge theories\cite{r2,r3,ng1}. The renormalization of
   superfluid density by $F_1$ appears similarly in the GL action and the superfluid density is renormalized by the
   same factor $1+F_1\sim x$ (Eq.\ (\ref{sdensity})). For weakly coupled layers of two-dimensional superconductors
   with $x<<1$, the renormalization of phase stiffness but not amplitude stiffness in the GL action implies that the
   temperature $T_c$ which is determined by the BKT transition\cite{bkt}, and $T^*\sim T_M$ which is determined by
   the BCS mean-field transition, separates. The BKT transition is determined by $\rho_s$ and happens at a
   temperature\cite{bkt}
   \[
   kT_c\sim{\hbar^2\over4m^*}\rho_s\sim{\hbar^2\over4m^*}x\rho(0),  \]
   which is much lower than the mean-field transition temperature $T_M$. The amplitude of gap parameter is
   only weakly renormalized in this temperature regime because $\xi_A$ is unrenormalized.

     We notice, however that many properties of high-$T_c$ cuprates are not explained by our simple model of
   Fermi-liquid superconductor which includes only $F_0$ and $F_1$, for example, the temperature dependence of
   London penetration depth $\lambda$ at low temperature\cite{r2,r3} ($T\rightarrow0$) and the apparent narrowness
   of paraconductivity regime\cite{r2,r3,e1} ($\sim$ several $T_c$, see also Ref.\cite{e4})) in the pseudo-gap phase
   at $T\gtrsim T_c$. In the vortex liquid phase, the conductivity of the system is given in the two fluid picture by
   $\sigma=\sigma_L+\sigma_v$, where $\sigma_L$ is the quasi-particle (normal) conductivity and $\sigma_v$ is the
   vortex liquid conductivity. In the BS picture, $\sigma_v\sim\eta_v/n_v\sim({a_v\over\xi_P^2})\sigma_L/n_v$, where
   $n_v$ is the vortex density. Therefore,
   \[
   \sigma\sim\sigma_L(1+{a_v\over\xi_P^2n_v}) ,\]
   and $\sigma_v$ dominates as long as $\xi_P^2n_v<<1$. At $T>>T_c$,
   $\xi_P^2n_v\sim e^{-\epsilon_c/kT}$, where $\epsilon_c$ is the vortex core energy. In GL theory,
   $\epsilon_c\sim(\hbar^2/2m^*)(\alpha/\beta)\sim E_f>>T_M$ in the weak coupling limit and crossover to
   $\epsilon_c\sim T_M$ in the strong coupling limit\cite{r3}. Therefore, $\epsilon_c$ is as least of order of
   $T^*\sim T_M$ and $\xi_P^2n_v<<1$ at temperatures $T<<T^*$, meaning that paraconductivity should dominate over most
   of the pseudo-gap regime in this simple GL model.

    Summarizing, in this paper we study the effect of Landau interactions $F_0$ and $F_1$ on the Ginzburg-Landau
   action for superconductors. We find that $F_0$ renormalizes the parameter $\beta$ in GL action but do not lead to
   qualitative changes. The effects of $F_1$ on GL action is much more interesting. It renormalizes superfluid density and
   separate the length (energy) scale for amplitude and phase fluctuations. For $1+F_1<<1$ it provides a natural
   mechanism for separation of the mean-field transition temperature ($T_M\sim T^*$) and BKT transition temperature
   ($T_c$) and suggests that the pseudo-gap phenomenon observed in high-$T_c$ cuprates may be a rather general
   property of Fermi-liquid superconductors with strong current renormalization $F_1<0$. We point out however that our simple
   model of Fermi liquid superconductor with Landau interactions $F_0$ and $F_1$ cannot explain many properties of the
   high-$T_c$ cuprates and a much more sophisticated model is needed to describe realistic underdoped cuprates.

  \acknowledgements
  This work is supported by HKUGC through grant number 602705.


\references
\bibitem{r1} A. Damascelli and Z. Hussain and Z. X. Shen, Rev. Mod. Phys. \textbf{75} 473 (2003).
\bibitem{r2} J. Orenstein and A.J. Millis, Science {\bf 288}, 468(2000).
\bibitem{r3} P. A. Lee and N. Nagaosa and X. G. Wen, Rev. Mod. Phys. \textbf{78}, 17 (2006).
\bibitem{emery} V.J. Emery and S. Kivelson, {\em Nature} {\bf 374}, 434(1995).
\bibitem{rand} C.A.R. S$\grave{a}$ de Melo, M. Randeria and J.R. Engelbrecht, \prl {\bf 71}, 3202 (1993).
\bibitem{levin} see for example, J. Stajic, A. Iyengar, Q. Chen, and K. Levin, \prb {\bf 68}, 174517 (2003).
\bibitem{e1} J. Corson, R. Mallozzi, J. Orenstein, J.N. Eckstein and I. Bozovic, {\em Nature} {\bf 398}, 221 (1999).
\bibitem{millis} L.B. Ioffe and A.J. Millis, \prb {\bf 66}, 094513 (2002)
\bibitem{leggett} A.J. Leggett, Phys. Rev. {\bf 140}, A1869(1965); A.I. Larkin, Sov. Phys.
 JETP {\bf 14}, 1498(1964).
\bibitem{tinkham} M. Tinkham, "Introduction to Superconductivity",(2nd, ed, McGraw Hill, New York, 1996).
\bibitem{ng1} T.K. Ng, \prb {\bf 69}, 125112(2004).
\bibitem{bkt} V.L. Berezinskii, Zh. Eksp. Teor, Fiz. {\bf 61} 1144 (1971); J.M. Kosterlitz and D.J. Thouless,
 J. Phys. {bf C 5}, L124 (1972).
\bibitem{e4} N.P. Ong, Yayu Wang, S. Ono, Y. Ando and S. Uchida, Ann. Phys. (Leipzig) {\bf 13}, No.1-2,9-14 (2004).
\end{document}